\documentclass[preprint2]{aastex}
\shorttitle{Catastrophic flux rope eruption}

\shortauthors{Chen, Hu {\&} Sun}

\begin{document}

\title{Catastrophic eruption of magnetic flux rope in the corona and
solar wind with and without magnetic reconnection}

\author{Y. Chen\altaffilmark{1,2}, Y. Q. Hu\altaffilmark{2}, AND S. J. SUN\altaffilmark{2}}

\altaffiltext{1}{Institute for Space Sciences, Shandong
University; Department of Space Science and Applied Physics,
Shandong University at Weihai, Weihai Shandong, 264209 China;
yaochen@sdu.edu.cn} \altaffiltext{2}{School of Earth and Space
Sciences, University of Science and Technology of China, Hefei
Anhui 230026, China}

\begin{abstract}
It is generally believed that the magnetic free energy accumulated
in the corona serves as a main energy source for solar explosions
such as coronal mass ejections (CMEs). In the framework of the
flux rope catastrophe model for CMEs, the energy may be abruptly
released either by an ideal magnetohydrodynamic (MHD) catastrophe,
which belongs to a global magnetic topological instability of the
system, or by a fast magnetic reconnection across preexisting or
rapidly-developing electric current sheets. Both ways of magnetic
energy release are thought to be important to CME dynamics. To
disentangle their contributions, we construct a flux rope
catastrophe model in the corona and solar wind and compare
different cases in which we either prohibit or allow magnetic
reconnection to take place across rapidly-growing current sheets
during the eruption. It is demonstrated that CMEs, even fast ones,
can be produced taking the ideal MHD catastrophe as the only
process of magnetic energy release. Nevertheless, the eruptive
speed can be significantly enhanced after magnetic reconnection
sets in. In addition, a smooth transition from slow to fast
eruptions is observed when increasing the strength of the
background magnetic field, simply because in a stronger field
there is more free magnetic energy at the catastrophic point
available to be released during an eruption. This suggests that
fast and slow CMEs may have an identical driving mechanism.
\end{abstract}


\section{INTRODUCTION}

It is generally believed that the accumulated magnetic free energy
serves as a main energy source for the spectacular solar eruptive
phenomena such as coronal mass ejections (CMEs) (see reviews by
Forbes (2000) and Low (2001)), but it remains open how the
magnetic energy is released. Among various scenarios, the flux
rope catastrophe mechanism is a very promising one (reviewed
recently by Lin et al. (2003) and Hu (2005)). The catastrophe is
an ideal magnetohydrodynamic (MHD) process belonging to a global
magnetic topological instability of the system. It releases energy
without ohmic heating, especially suitable for CMEs without
associated flares (e.g., Forbes {\&} Isenberg, 1991; Isenberg et
al., 1993; Forbes {\&} Priest, 1995; Hu et al., 2003b). A
by-product of the catastrophe is the formation of one or more
electric current sheets, which grows at the Alfv\'enic time scale
and provides proper sites for fast magnetic reconnection. Such a
reconnection further releases the stored magnetic energy and
should be responsible for a solar flare associated with a CME
event (e.g., Forbes {\&} Lin, 2000; Lin {\&} Forbes, 2000). Thus,
in terms of this scenario, there are two means of magnetic energy
release process involved in solar eruptions, which are the ideal
MHD catastrophe and resistive magnetic reconnection. Both
processes are thought to be important to the CME initiation and
acceleration. It is therefore necessary and important to
disentangle their contributions to the CME dynamics and
energetics. This article serves as a first step to solve this
problem under the framework of the specific catastrophe model for
CMEs.

The eruptive flux rope after catastrophe is accelerated mainly by
the net or unbalanced Lorentz force, which is contributed by
various source currents both inside and outside the flux rope,
including that along each electric current sheet and the
corresponding image current, and the background potential field
which is determined by the normal component of the background
field at the solar surface. To reveal the roles of these various
components of magnetic forces in sustaining the flux rope in
equilibrium and causing the eruption, Chen et al. (2006b)
conducted detailed force balance analyses. It was found that the
primary lifting force is provided by the azimuthal current inside
the rope and its image below the photosphere, which is mainly
balanced by the pulling force produced by the background potential
field when the rope is in equilibrium. During an eruption of the
rope caused by the catastrophe, the force associated with the
rapidly-developing current sheet(s) constitutes an additional and
significant restoring force that decelerates the rope. This force
will be greatly reduced or even eliminated once magnetic
reconnection sets in across the current sheet(s), as a consequence
the magnetic reconnection may cause a further acceleration of the
flux rope. This will be confirmed and quantified by the
calculations presented in this article with an axisymmetric flux
rope catastrophe model. The effect of the solar wind plasma with
an equatorial current sheet above the streamer cusp point
extending to infinity is considered so as to get a more realistic
description of the CME acceleration and propagation. Note that in
most previous studies on the flux rope catastrophe this effect was
not included. Sun {\&} Hu (2005) did consider the effect of the
solar wind on the rope catastrophe and the consequent eruption.
They suggested that the flux rope catastrophe can serve as a
mechanism for slow CMEs. Yet, they investigated only the cases
corresponding to a relatively weak background field in terms of
ideal MHD. Based on their work, we take the effect of both the
solar wind and magnetic reconnection into account in the present
study, which may be regarded as a starting point to disentangle
the impacts of MHD catastrophe and magnetic reconnection on the
CME dynamics in the solar wind. We put our focus on the
differences between the flux rope dynamics in two situations in
which we either prohibit or allow magnetic reconnection to take
place across the rapidly-growing current sheets. We introduce the
background corona and solar wind model and the flux rope
catastrophe model in the following section. In $\S$ 3.1 we compare
the solutions with and without magnetic reconnection to illustrate
its impact on the CME dynamics. We emphasize the effect of the
background magnetic field strength in $\S$ 3.2 and provide a brief
summary of this article with discussions in the last section of
this article.

\section{FLUX ROPE CATASTROPHE MODEL IN THE CORONA AND SOLAR WIND}
Coronal magnetic flux ropes are in essence three-dimensional
structures with two ends anchored to the photosphere. Considering
that the length of a magnetic flux rope is much larger than its
diameter, the axisymmetrical simplification (also called 2.5-D
models) can be used to approximate the realistic flux rope system.
With this approximation, the magnetic field $\mathbf{B}$ can be
expressed with a magnetic flux function $\psi(t,r,\theta)$ in
spherical coordinates ($r$, $\theta$, $\varphi$),
\begin{equation}
\emph{\textbf{B}}= \nabla\times({\psi \over
r\sin\theta}\mathbf{\hat{\varphi}})
+B_\varphi\mathbf{\hat{\varphi}}. %\eqno{(1)}
\end{equation}
The derived resistive MHD equations in spherical coordinates are
written into the following form (see also Ding et al., 2006):
\begin{equation}
{\partial \rho \over \partial
t}+\nabla\cdot(\rho\emph{\textbf{v}})=0, %\eqno{(2)}
\end{equation}
\begin{equation}
\begin{array}{c}
{\partial \emph{\textbf{v}} \over \partial t} +
\emph{\textbf{v}}\cdot\nabla\emph{\textbf{v}}+{1\over \rho}\nabla
p+{1\over \mu\rho}[L\psi\nabla\psi+
\emph{\textbf{B}}_\varphi\times(\nabla\times\emph{\textbf{B}}_\varphi)]\\[3mm]
+{1\over\mu\rho
r\sin{\theta}}\nabla\psi\cdot(\nabla\times\emph{\textbf{B}}_\varphi)\hat{\varphi}+{GM_\odot\over
r^2}\hat{r}=0, %\eqno{(3)}
\end{array}
\end{equation}
\begin{equation}
{\partial \psi\over\partial
t}+\emph{\textbf{v}}\cdot\nabla\psi-{1\over\mu}\eta
r^2\sin^2\theta L\psi=0, %\eqno{(4)}
\end{equation}
\begin{equation}
\begin{array}{c}
{\partial B_\varphi \over \partial t}
+r\sin\theta\nabla\cdot({B_\varphi\emph{\textbf{v}}\over
r\sin\theta})+[\nabla\psi\times\nabla({v_\varphi\over r
\sin\theta})]_\varphi\\[3mm]
-{1\over r\sin\theta}\nabla\eta \cdot\nabla(\mu r\sin\theta
B_\varphi)-{1\over\mu}\eta r \sin\theta L(r
B_\varphi\sin\theta)=0,
% \eqno{(5)}
\end{array}
\end{equation}
\begin{equation}
{\partial T\over \partial t}+\emph{\textbf{v}}\cdot\nabla
T+(\gamma-1)T\nabla\cdot \emph{\textbf{v}}-{{\gamma-1}\over \rho
R}\eta \emph{\textbf{j}}^2=0, %\eqno{(6)}
\end{equation}
where
\begin{equation}
L\equiv{1\over r^2\sin^2\theta}({\partial^2 \over
\partial r^2}+{1\over r^2}{\partial^2 \over \partial \theta^2}
-{\cot\theta \over r^2} {\partial \over \partial \theta}),
%\eqno{(7)}
\end{equation}
\begin{equation}
\emph{\textbf{j}}={1\over \mu}\nabla\times
\emph{\textbf{B}}=-{1\over \mu}r\sin\theta L\psi\hat\varphi +
{1\over\mu}\nabla\times(B_\varphi\hat\varphi). %\eqno{(8)}
\end{equation}
The symbols $\rho$, $\emph{\textbf{v}}$, $T$, and
$\emph{\textbf{j}}$ represent the density, flow velocity,
temperature and the current density, respectively. $\mu$ is the
vacuum magnetic permeability, $R$ the gas constant, $G$ the
gravitational constant, $M_\odot$ the solar mass, and $\gamma$ the
polytropic index which is taken to be 1.05 so as to obtain a
reasonable solution of the background steady-state solar wind. The
temperature and density at the coronal base are taken to be
$T_0=2\times10^6$ K and $\rho_0=1.67\times10^{-13}$ kg m$^{-3}$,
respectively. The magnetic flux function at the coronal base is
taken to be the same as a dipolar field given by
$\psi(r=R_\odot)=\psi_0 \sin^2\theta/R_\odot$ where $\psi_0=B_0
{R_\odot}^2$. $B_0$ represents half of the magnetic field strength
at the polar hole ($\theta=0$) on the solar surface, which can be
used to represent the strength of the background magnetic field of
the system. Note that in this study the magnetic flux function at
the solar surface is fixed to its initial distribution which is
determined by $B_0$. For the effect of the photospheric magnetic
flux distribution on the coronal flux rope catastrophe, please
check a recent paper written by Sun et al. (2007). The parameter
$B_0$ will be freely adjusted. For example, $\psi_0=9.7 \times
10^{13}$ Wb with $B_0$ taken to be 2 G giving the case
investigated by Sun {\&} Hu (2005). It is defined that the ideal
MHD situation corresponds to a zero magnetic resistivity $\eta$.
In the resistive situation, an anomalous homogeneous resistivity
is used with $\eta$ given by $\eta=\eta_0 \mu v_{s0}R_\odot$ where
$\eta_0=0.1$ and $v_{s0}=\sqrt{2RT_0}=181.8$ km s$^{-1}$. The
steady-state polytropic solar wind solution is obtained by solving
the above MHD equations with $\eta=0$. The magnetic topology
(white lines) and the velocity color contour map for $B_0=6$ G
from 1 $R_\odot$ to 20 $R_\odot$ are illustrated in Figure 1a. For
other values of $B_0$, the field topology and velocity
distribution are basically similar to that shown in this figure.
It can be seen that the solution is characterized by a typical
streamer-current sheet-solar wind configuration. The cusp point is
located at about 3 $R_\odot$, and the flow velocity reaches up to
400 km s$^{-1}$ at about 10 $R_\odot$ along the equator.

Based on the obtained corona and solar wind solution, we let a
flux rope with prescribed mass, toroidal and poloidal magnetic
fluxes ($\Psi_p$ and $\Psi_\varphi$) emerge from the equator at
the coronal base. The detailed emerging process has been described
previously by Hu et al. (2003b) and Chen et al. (2006a) and will
not be repeated here. Special numerical measures are taken to
maintain $\Psi_p$ and $\Psi_\varphi$ invariant and equal to their
initial given values during the simulation (see Hu et al., 2003b).
Figure 1b exemplifies the magnetic configuration and velocity
contour map of a flux-rope system with the solar wind, where the
border of the original flux rope is depicted with a green circle.
It can be seen that the closed field region of the streamer
expands apparently with the emergence of the flux rope. Such a
swelling of a coronal streamer is often observed with the
white-light coronagraphs before CMEs (e.g., Howard et al., 1985).
The mass contained by the flux rope per radian in the azimuthal
direction is set to be ${0.5 \over 2 \pi} M_0$ where the unit of
mass $M_0=\rho_o R_\odot^3=5.643\times10^{13}$ kg, the poloidal
flux $\Psi_p$ is taken to be 0.3 in units of $\psi_0$ while the
toroidal flux $\Psi_\varphi$ is changeable. Thus one may find MHD
solutions with different values of $\Psi_\varphi$ to examine
whether a catastrophe occurs, and find out the meta-stable state
of the system characterized by $\Psi_p=0.3$ and a specific value
of $\Psi_\varphi$ which depends on $B_0$. Starting from this state
any slight increase of $\Psi_p$ or $\Psi_\varphi$ may excite the
catastrophe. Therefore, the state is taken as the initial state
for our simulation of the flux rope eruption. We choose to
increase $\Psi_p$ from $0.3$ to $0.305$ at $t=0$ so as to trigger
the catastrophe. Physically speaking, the increase of the poloidal
flux can be achieved by a twist of a long three-dimensional flux
rope anchored to the photosphere. Besides the mass, magnetic
fluxes of the flux rope, the helicity of the flux rope is also of
interest to the study on CMEs. In an axisymmetric system like that
investigated in this article, the two-dimensional magnetic
helicity can be calculated by the following integral according to
Hu et al. (1997),
$$
H_T=2\pi\int\int \psi B_\varphi r dr d\theta, \eqno{(9)}$$ where
the factor $2\pi$ comes from the integral over the azimuthal
direction, which should be removed if one wants to evaluate the
magnetic helicity per radian in the azimuthal direction.

When the flux rope breaks away from the surface and erupts
upwards, a current sheet may develop below the flux rope. It is
well known that numerical pseudoreconnection takes place across
the current sheet in most numerical simulations, which causes a
false transfer of poloidal flux from the background to the flux
rope and results in a topological change. In this work, we take
special measure to prohibit such numerical reconnection in order
to investigate the flux rope dynamics in the framework of ideal
MHD. The magnetic flux function $\psi$ along the current sheet is
invariant, which is known a priori, and any reconnection across
the sheet reduces it in the present simulation. We therefore
reassign $\psi$ along the current sheet to the known constant
value at each time step. This technique, first proposed by Hu et
al. (2003b), effectively eliminates numerical reconnection across
the equatorial current sheet. Note that this special measure is
not employed in our calculations in terms of resistive MHD with a
non-zero $\eta$.

The calculations are carried out in a domain of $R_\odot \le r \le
30 R_\odot$ and $0 \le \theta \le \pi/2$, which is discretized
into 150 $\times$ 90 grid points. The grid spacing increases
according to a geometric series of a common ratio 1.024 along the
radial direction from 0.02 at the solar surface to 0.71 at the top
boundary. And a uniform mesh is adopted in the $\theta$ direction.
The multistep implicit scheme developed by Hu (1989) is used to
solve the MHD equations. For the eruptive solutions, the
calculations are terminated once the top part of the ejecta
reaches the upper boundary.

\section{Numerical results}
In this section, we first present and compare the solutions given
by the ideal and resistive MHD calculations for the case with
$B_0=6$ G as a first step to disentangle the impacts of MHD
catastrophe and magnetic reconnection on the CME dynamics in the
solar wind. Then, we investigate the effect of the background
magnetic field strength by comparing results with different values
of $B_0$.

\subsection{Impact of magnetic reconnection on flux rope dynamics}
As mentioned previously, our simulations on the flux rope
catastrophe and eruption start from an equilibrium state which is
a meta-stable flux rope system in the corona and solar wind
background. The catastrophe is triggered by a slight increase of
the poloidal flux $\Psi_p$ of the flux rope from 0.3 to 0.305 (or
8.7 - 8.85 $\times 10^{13}$ Wb in physical units) at $t=0$ with
the critical axial flux in the rope $\Psi_\varphi=0.209$ (6.06
$\times 10^{13}$ Wb) for $B_0$=6 G. After that, the flux rope
starts to break away from the photosphere and erupts upwards.
Figures 1c and 1d show the magnetic topology and velocity color
contours at the same instant ($t=280$ minutes) for the two
solutions with and without magnetic reconnection (i.e., the cases
with $\eta \ne 0$ and $\eta = 0$). An apparent difference between
the two solutions lies in whether a current sheet develops below
the flux rope. The sheet forms and grows with the rope eruption in
the ideal MHD case, while it is eroded by magnetic reconnection in
the resistive MHD calculation. It is also apparent that a
significant part of the poloidal flux has been transferred from
the background to the flux rope, and a new streamer appears as the
aftermath of magnetic reconnection in the resistive calculation.
The rate of magnetic flux transfer is mainly determined by the
effective resistivity consisting of the anomalous resistivity and
the numerical resistivity involved in the pseudoreconnection.
Unfortunately, at this time, it is not possible to eliminate the
pseudoreconnection in our calculations in terms of resistive MHD.
Therefore, it is difficult to control the flux transfer rate by
simply adjusting the magnetic resistivity in this case. Further
discussion regarding this issue will be given in our discussion
section. Another major difference is the color distribution of the
velocity contour maps which indicates how fast the flux rope
ejecta is. It can be seen that the rope erupts faster in the
solution with magnetic reconnection, as will be quantitatively
revealed in Figure 2.

In Figure 2, we plot the profiles of the heliocentric distance,
velocity, and acceleration of different parts of the flux rope
ejecta, including the cusp point (in dotted), the rope top (in
dashed), the rope axis (in solid), and the rope bottom (in
dot-dashed), left panels for the case without reconnection and
right for the reconnection case. It can be seen that in both cases
the flux rope starts to take off at $t \approx 70$ minutes. There
is an apparent delay of about 50 minutes of the time when the cusp
point starts to move upwards rapidly. The delay reflects the time
taken for the eruptive flux rope to propagate from the coronal
base to the initial cusp point location in the corona. We can see
that the cusp point undergoes the most dramatic acceleration in
both cases: in about 70 minutes the velocity of the cusp point
reaches up to 800 km s$^{-1}$ with the maximum acceleration being
300 m s$^{-2}$ in the ideal case, and 1200 km s$^{-1}$ and 500 m
s$^{-2}$ for the resistive one. On the other hand, it takes 2 to 3
hours for the flux rope to be accelerated to the maximum speed.
After the maximum, the velocities become more or less constant.
The velocities of different parts of the ejecta vary significantly
from the cusp point to the rope bottom. For example, at $t=280$
minutes the exact moment at which the snapshot is taken for Figure
1, the velocity decreases monotonically from about 800 km s$^{-1}$
at the cusp point to 350 km s$^{-1}$ at the rope bottom in the
ideal case, and from 1100 km s$^{-1}$ to 750 km s$^{-1}$ in the
resistive case. This monotonic decrease of velocities from the
leading to trailing edges has been often observed by measurements
of flux-rope like structures in the interplanetary space (e.g.,
Gosling et al., 1998), which simply indicates that the rope
undergoes a rapid expansion during its eruption. The velocity
profiles given by both solutions in Figure 2 as well as that in
the following figure are in a good agreement with a recent
statistical study on the CME accelerations, which indicates that a
CME usually undergoes multiphased kinematic evolution including an
initial slow rise phase and a main rapid acceleration phase in the
inner corona, and a relatively smooth propagation phase in the
outer corona (Zhang {\&} Dere, 2006). The magnitude and duration
of the main acceleration given by our calculations are also in
line with their statistical results. Taking the resistive case
shown in Figure 2 as an example, the main acceleration phase
starts from $t\approx 70$ minutes and ends at about $t=190$
minutes lasting for nearly 2 hours. Comparing the solutions with
and without magnetic reconnection, it is apparent that the speeds
and accelerations of the flux rope are significantly enhanced in
the case involved with magnetic reconnection. Further discussions
regarding the roles of magnetic reconnection in the CME dynamics
will be given in our discussion section.

\subsection{Effect of background field strength}
In this subsection, we present numerical results given by
calculations with different values of the parameter $B_0$, which
represents the strength of the background field and directly
relates to the amounts of magnetic energy that can be stored and
released in the system. In Figure 3, we plot radial profiles of
the velocity and acceleration of different parts of the system
including the cusp point (in dotted), and the rope top (in
dashed), axis (in solid), and bottom (in dot-dashed) for the cases
with (thick lines) and without (thin lines) reconnection, the left
panels are for the solution with $B_0=2$ G and the right panels
with $B_0=10$ G. Note that for clearness only the accelerations of
the cusp point and rope axis are plotted in the lower panels.
Similar to the above calculation the catastrophe is triggered by a
slight increase of the dimensionless poloidal flux inside the rope
$\Psi_p$ from 0.3 to 0.305 with the corresponding critical
dimensionless axial flux $\Psi_\varphi$ equal to 0.129 for the
case with $B_0=2$ G and 0.244 for $B_0=10$ G. Note that the
magnetic fluxes in physical units are listed in Table 1. Since the
relative magnitude of the rope poloidal flux does not change with
varying $B_0$, the size of the flux rope does not change
apparently either. To be quantitatively, we checked the
heliocentric distance of the rope axis $r_{a}$ which can be used
to represent the size of the flux rope. It was found that
$r_a$=1.42, 1.52, and 1.54 solar radii for the cases with $B_0$=2,
6, and 10 G, respectively. As a result, the plasma density inside
the flux rope gets slightly different in different cases since the
total mass contained by the rope does not change with $B_0$. In
Table 1, we also list the two-dimensional magnetic helicity per
radian in the azimuthal direction $H_T$ calculated with Equation
(9). Similar as the rope fluxes, $H_T$ also increases dramatically
with increasing $B_0$. We see that the physical properties of the
flux rope differ significantly from case to case in our
calculations, which naturally have impacts on the flux rope
dynamics. Further discussion along this direction will be
presented at the end of the following paragraph.

It can be seen from Figure 3 that the most obvious difference
between the two sets of solution is the magnitude of velocity and
acceleration at different points of the ejecta. In the resistive
calculation with $B_0=2$ G, the main acceleration phase of the
flux rope lasts for about 4 hours from $t\approx 100$ minutes to
$t\approx 350$ minutes with speeds rising up to 670 km s$^{-1}$ at
the cusp point and to 460 km s$^{-1}$ at the rope bottom, which is
followed by the so-called propagation phase with a nearly constant
speed (Zhang {\&} Dere, 2006). In the ideal calculation with the
same value of $B_0$, the velocity keeps increasing till $t=600$
minutes, yet the corresponding acceleration gets smaller than 5 m
s$^{-2}$. In the resistive case with $B_0=10$ G, the main
acceleration phase lasts for about 2 hours starting from $t\approx
40$ minutes and ending at about $t=150$ minutes with the maximum
acceleration rising up to 750 m s$^{-2}$. After the maximum of
velocity is reached, which is about 1600 km s$^{-1}$ for the cusp
point and 1150 km s$^{-1}$ for the rope bottom, the flux rope gets
decelerated gradually to a velocity of 1300 km s$^{-1}$ at the
cusp point and 900 km s$^{-1}$ at the rope bottom. In the ideal
case with $B_0=10$ G, the main acceleration phase also lasts for
about 2 hours with the maximum acceleration being about 430 m
s$^{-2}$ for the cusp point and 220 m s$^{-2}$ for the rope axis.
In the propagation phase following the velocity maximum, the
velocities decrease slightly. It can be seen that CMEs, even fast
ones, can be produced taking the ideal MHD catastrophe as the only
process of magnetic energy release. It is also true, again, that
the eruptive speeds are significantly enhanced after magnetic
reconnection sets in. We point it out in passing that a smooth
transition of eruptions from slow to fast can be obtained when
varying $B_0$ continuously with a stronger background field
corresponding to a faster eruption, in line with the very recent
study on the effect of photospheric flux distribution on the flux
rope dynamics by Sun et al. (2007). The physical cause of such a
behavior can be easily understood from the following simple energy
analysis. Since the pattern of the magnetic flux distribution at
the coronal base remains the same for all cases we have discussed,
the associated open field energy must be proportional to the
square of $B_0$. On the other hand, the percentage by which the
catastrophic energy threshold exceeds the open field energy varies
in a much smaller range. It reads 6.1{\%}, 8.8\% and 9.6\% for
$B_0$ = 2 G, 6 G and 10 G, respectively. Such a result is
consistent with previous similar calculations (e.g., Hu et al.,
2003b, Li {\&} Hu, 2003; Chen et al., 2006a). Therefore, the total
ammount of magnetic free energy of the system at the catastrophic
point is mainly determined by the overall strength of the
background field in spite of the dramatic differences in the flux
rope properties at the catastrophic point as listed in Table 1. In
summary, we argue that the stronger the background field is, the
more magnetic free energy is available for the flux rope system at
the catastrophic point, which leads to a faster eruption of the
flux rope.

To shed more light on the effect of magnetic reconnection on the
CME dynamics, we calculate the total increase in kinetic energy of
the system compared with that of the pre-eruption state (i.e., the
state at $t=0$), represented by $\Delta E_k$. Figure 4 shows the
temporal profiles of $\Delta E_k$ per radian in the azimuthal
direction in units of $5.38\times 10^{31}$ ergs for the three sets
of solutions with $B_0$=2 G (in solid), 6 G (in dotted), and 10 G
(in dashed), where thick and thin lines represent the results with
and without magnetic reconnection. The velocity and acceleration
for these solutions have been illustrated in Figures 2 and 3. We
can see that $\Delta E_k$ tends to reach an asymptotic value in
all cases. For each set of solutions, the asymptotic value of
$\Delta E_k$ in the resistive case is about 2 to 3 times of that
in the ideal case indicating that the MHD catastrophe and magnetic
reconnection, the two means of magnetic energy release process,
are of comparable importance on the CME acceleration for the
resistive MHD situation.

\section{Conclusions and discussion}
In terms of the catastrophe theory, there are two main processes
energizing the solar eruptions: MHD catastrophe and magnetic
reconnection. This article serves as a first step to disentangle
their contributions to the CME dynamics. To do this, we construct
a flux rope catastrophe model in the corona and solar wind and
compare different cases in which we either prohibit or allow
magnetic reconnection to take place across rapidly-growing current
sheets during the eruption. For simplicity, a polytropic process
with the polytropic index $\gamma=$1.05 is used to produce the
background solar wind solution. The catastrophe and the consequent
eruption is triggered by a tiny increase of the rope poloidal
flux, which reflects a slight twist of the ends of a long
three-dimensional realistic flux rope anchored to the photosphere.
It is demonstrated that CMEs, even fast ones, can be produced
taking the ideal MHD catastrophe as the only process of magnetic
energy release. Nevertheless, the eruptive speed can be
significantly enhanced after magnetic reconnection sets in. In
addition, a smooth transition from slow to fast eruptions is
yielded when increasing the strength of the background magnetic
field, i.e., a stronger field, in which more free magnetic energy
gets available at the catastrophic point, enables a faster
eruption. This suggests that fast and slow CMEs may have an
identical driving mechanism.

Based on previous and present studies taking catastrophe as the
principle driving mechanism of CMEs, we argue that the MHD
catastrophe is probably the main means of energy release for CMEs
at least in the initial phase. It releases energy without ohmic
heating and provide accelerations with the Lorentz force,
especially suitable for non-flare associated CMEs. A by-product of
the catastrophe is the formation of one or more electric current
sheets, which proceeds at the Alfv\'enic time scale and produces
conditions favoring fast magnetic reconnection. Such a
reconnection, if takes place, further releases the magnetic energy
through the following two aspects. Firstly, the magnetic energy is
converted into thermal and kinetic energies of plasma particles at
the reconnection site. This process is believed to account for a
solar flare associated with a CME event. Secondly, the restoring
force contributed by the current in the current sheet is
significantly reduced or even eliminated, and the magnetic
topology changes with the magnetic reconnection. This also
produces a significant acceleration of the flux rope in addition
to that caused by catastrophe. It can be seen from our
quantitative calculation that the MHD catastrophe and magnetic
reconnection, the two magnetic energy release processes, may have
comparable impacts on the CME dynamics during the main
acceleration phase of CMEs.

A major subject of this article is to estimate the impact of
magnetic reconnection on the flux rope dynamics. For this purpose,
we compared solutions given by calculations with and without
magnetic reconnection. However, this work suffers from the facts
that the anomalous magnetic resistivity is artificially given and
the numerical pseudoreconnection is unavoidable in the resistive
calculations. Since the effective resistivity including both the
anomalous and the numerical ones is believed to be much larger
than the realistic value in the corona and solar wind, the
velocity profiles given by our study for the case with magnetic
reconnection should be taken as the upper bound for the realistic
situation. As mentioned in the text, the effective resistivity is
a crucial factor determining the transfer rate of magnetic flux
from the background to the flux rope. Observationally, this
transfer rate can be evaluated by extrapolating the photospheric
field to the corona and counting the change of the total magnetic
flux in the aftermath of a CME event, e.g., in the coronal dimming
region (see, e.g., Jing et al., 2005 and Qiu {\&} Yurchyshyn,
2005). Further theoretical endeavor should utilize these relevant
observational constraints on the flux transfer rate in the
modelling of a specific event.

Our study on the effect of magnetic field strength reveals a
smooth transition from slow to fast eruptions when increasing the
background field strength. This is in support of the argument that
slow and fast CMEs may be driven by a single identical mechanism,
also in line with recent statistical results contradicting with
the traditional bimodal classification of slow (gradual) and fast
(impulsive) CMEs (e.g., Sheeley et al., 1999; Andrews {\&} Howard,
2001). For instance, it is shown by Yurchyshyn et al. (2005) and
Zhang {\&} Dere (2006) that the velocity and acceleration of a
large amount of CME events have a continuous distribution instead
of a bimodal one, and by Vr$\breve{s}$nak et al. (2005) that
flare- and nonflare- associated CMEs have quite similar
characteristics in the LASCO C2 and C3 fields of view. The
velocity profiles for slow CMEs given by our model, say, the
solution corresponding to a weak background field, show a gradual
acceleration, while that for fast CMEs present a rapid
acceleration and a discernable deceleration following the main
acceleration phase. This behavior is probably a result of the
coupling process with the background solar wind plasma according
to our preliminary analysis. When the magnetic energy released in
an eruption is not enough to accelerate the flux rope ejecta to
the speed of the background plasmas, the ejecta may get gradually
accelerated and gain more energy through the coupling to the solar
wind. On the other hand, the ejecta may get decelerated and lose
energy through similar coupling process with the solar wind while
the released magnetic energy is enough to push the flux rope
outwards with a velocity faster than that of the background.

There exist contradicting discrepancies between the present
thick-rope model in axisymmetrical spherical geometry and that
published in the literature in terms of thin-rope models in 2-D
cartesian geometry. Firstly, an infinite amount of energy is
required to open up a closed magnetic field in 2-D Cartesian
geometry (Hu et al., 2003b), therefore, it is energetically
impossible to open the overlying field and to let the flux rope
escape to infinity without magnetic reconnection, as demonstrated
by the catastrophe models assuming Cartesian geometry (e.g., Lin
{\&} Forbes, 2000). On the other hand, in the spherical geometry
the corresponding open-field energy is finite and it can be
exceeded by the flux rope system as already shown by many
calculations (e.g., Choe {\&} Cheng, 2002; Hu et al., 2003b; Li
{\&} Hu, 2003; Flyer et al., 2004; Sun {\&} Hu, 2005; Zhang et
al., 2005; Peng {\&} Hu, 2005; Ding {\&} Hu, 2006; Chen et al.,
2006a). Thus, magnetic reconnection may not be necessarily
required for the flux rope to get escaped from the Sun in the
spherical model. Secondly, as pointed out previously, the
self-interaction of the azimuthal current inside the flux rope by
itself results in an outward radial force on the rope, which comes
from the curvature of the rope surrounding the Sun (Chen, 1989;
Lin et al., 1998; Krall et al., 2000; Chen et al., 2006b). This
self-force, together with that contributed by its image current
below the photosphere, serves as a dominant driving force for the
rope eruption. However, in the 2-D Cartesian models this
self-force is trivially zero by the symmetry of an infinitely long
straight current, this gives another basic difference between the
2-D Cartesian and spherical models.

No matter what geometry is used, so far most flux rope models have
been limited to 2-D analyses, as a necessary simplification for
practical tractability. Yet, the two ends of a 3-D flux rope are
believed to be anchored to the solar surface. It remains open
regarding how the catastrophic behavior of the flux rope may
change under this situation. Finally, since how the corona and
solar wind plasma is heated and accelerated is still a big issue
to be resolved, the polytropic process is assumed conveniently to
obtain the reasonable background solar wind solution in this work.
However, it should be noted that the polytropic solar wind
solution is too simple to account for some realistic properties of
the solar wind. For example, the effect of the fast solar wind is
not included in this study. This will certainly affect the
propagation of the ejecta at high latitudes, yet may not be very
important to the study on the CME propagation along the equatorial
plane. In future we consider to employ sophisticated heating
functions to produce a more realistic solar wind background (see,
e.g., Chen {\&}Hu, 2001; Hu et al., 2003a) for a more elaborated
study on the CME propagation in the meridional plane.

\acknowledgements
This work was supported by grants NNSFC 40404013, NSBRSF
G2006CB806304, and NNSFC 10233050 in China.

Table 1.\hspace{1em}The poloidal and toroidal fluxes ($\Psi_{pc}$
and $\Psi_{\varphi c}$), and the two-dimensional magnetic helicity
$H_T$ (per radian in the azimuthal direction) of the flux rope
system at the catastrophic point in different cases. \vskip 10pt

\begin{center}
\begin{tabular}{lccc}
$B_0$ (G) & 2 & 6 & 10 \\
\hline\hline
$\Psi_{pc}$ (Wb) & 2.91 $\times 10^{13}$ & 8.70 $\times 10^{13}$ & 1.45 $\times 10^{14}$\\
$\Psi_{\varphi c}$ (Wb) & 1.25 $\times 10^{13}$ & 6.06 $\times 10^{13}$ & 1.08 $\times 10^{14}$\\
$H_T$ (Wb$^2$) & 1.45 $\times 10^{27}$ & 2.12 $\times 10^{28}$ & 6.32 $\times 10^{28}$ \\
\hline
\end{tabular}
\end{center}

\begin{figure}
\epsscale{1.} \plotone{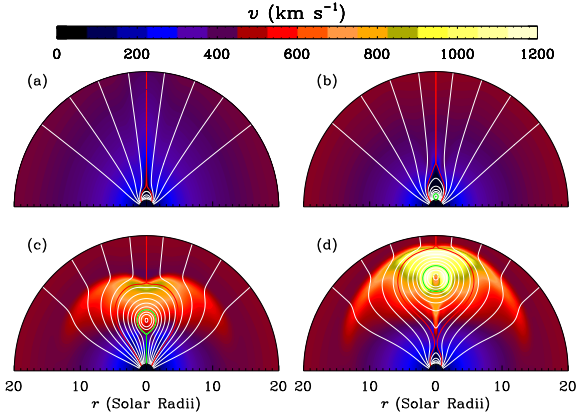} \caption{The magnetic topology
(white lines) and velocity color contour map for $B_0=6$ G from 1
$R_\odot$ to 20 $R_\odot$ for: (a) the background corona and solar
wind solution before the emergence of the flux rope, (b) the
pre-eruption state of the flux-rope system in the solar wind
background, (c) the solution with an erupting flux rope at $t=280$
minutes without magnetic reconnection, and (d) the eruptive
solution with magnetic reconnection at the same instant as panel
(c). The outer boundary of the original flux rope is depicted with
a green circle. \label{fig1}}
\end{figure}
\begin{figure}
\epsscale{1.} \plotone{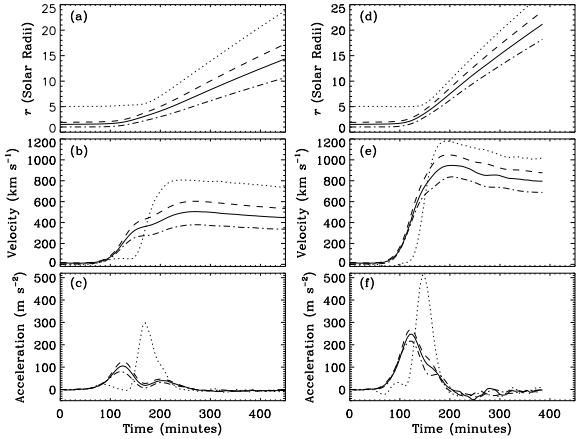} \caption{The temporal profiles of
the heliocentric distance, velocity, and acceleration of different
parts of the flux rope system, including the cusp point (in
dotted), the rope top (in dashed), the rope axis (in solid), and
the rope bottom (in dot-dashed). Left panels are for the case
without reconnection and right for the reconnection case.
\label{fig2}}
\end{figure}
\begin{figure}
\epsscale{1.} \plotone{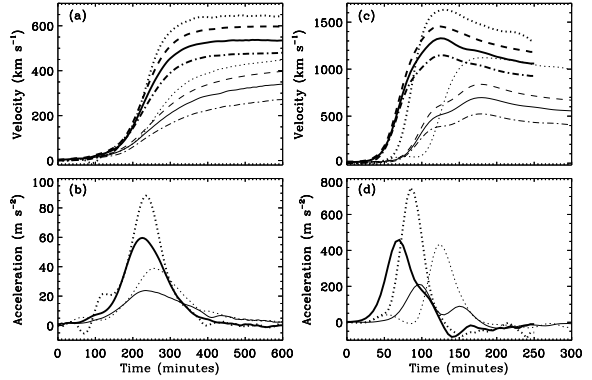} \caption{The radial profiles of the
velocity and acceleration of different parts of the system
including the cusp point (in dotted), and the rope top (in
dashed), axis (in solid), and bottom (in dot-dashed), the left
panels are for the case with $B_0=2$ G and the right panels with
$B_0=10$ G for the cases with (thick lines) and without (thin
lines) reconnection. \label{fig3}}
\end{figure}
\begin{figure}
\epsscale{1.} \plotone{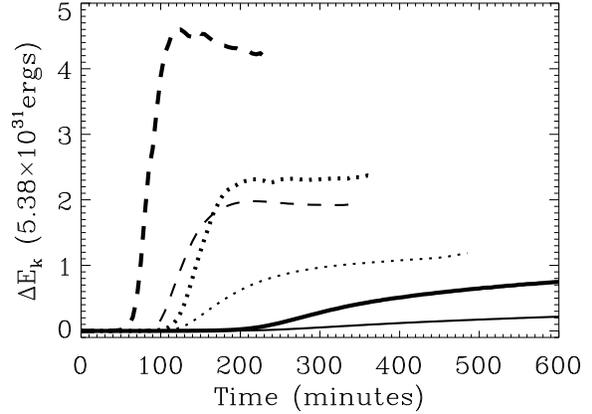} \caption{The temporal profiles of
the total increase in kinetic energy of the system $\Delta E_k$ in
units of $5.38\times 10^{31}$ ergs (per radian in the azimuthal
direction) for three sets of solutions ($B_0$=2 G (in solid), 6 G
(in dotted), and 10 G (in dashed)) with (thick lines) and without
(thin lines) magnetic reconnection. \label{fig4}}
\end{figure}


\begin{thebibliography}{}
\bibitem[Andrews 2001]{and01}
Andrews, M. D., \& Howard, R. S. 2001, Space Sci. Rev., 95, 147

\bibitem[Chen (1989)]{Chen89}
Chen, J., 1989, ApJ, 338, 453

\bibitem[Chen et al. 2006a]{che06_01}
Chen, Y., Chen, X. H., \& Hu, Y. Q. 2006a, ApJ, 644, 587

\bibitem[Chenhu 2001]{chen01}
Chen, Y., {\&} Hu, Y. Q. 2001, Sol. Phys., 199, 371

\bibitem[Chen et al. 2006b]{che06_02}
Chen, Y., Li, G. Q., \& Hu, Y. Q. 2006b, ApJ, 649, 1093

\bibitem[choe \& Cheng (2002)]{choe02}
Choe, G. S., \& Cheng C. Z., 2002, ApJ, 574, L179

\bibitem[Ding \& Hu 2006]{din06}
Ding, J. Y., \& Hu, Y. Q., 2006, Sol. Phys., 199, 371

\bibitem[Ding \& Hu 2006]{ding06}
Ding, J. Y., Hu, Y. Q., \& Wang, J. X., 2006, Sol. Phys., 235, 223

\bibitem[Flyer, Fornberg, Thomas, \& Low (2004)]{Flyer04}
Flyer, N., Fornberg, B., Thomas, S., \& Low, B. C., 2004, ApJ,
606, 1210

\bibitem[Forbes (2000)]{Forbes00_01}
Forbes, T. G., 2000, J. Geophys. Res., 105, 23153

\bibitem[Forbes \& Isenberg (1991)]{Forbes91}
Forbes, T. G., \& Isenberg, P. A., 1991, ApJ, 373, 294

\bibitem[Forbes \& Lin(2000)]{Forbes00_02}
Forbes, T. G., \& Lin, J., 2000, J. Atmos. Sol.-Terr. Phys., 62,
1499

\bibitem[Forbes \& Priest (1995)]{Forbes95}
Forbes, T. G., \& Priest, E. R., 1995, ApJ, 446, 377

\bibitem[Gosling 1998]{gos98}
Gosling, J. T., Riley, P., McComas, D. J., {\&} Pizzo, V. J.,
1998, J. Geophys. Res., 103, 1941

\bibitem[howard 1985]{how85}
Howard, R. A., Sheely, N. R., Jr., Koomen, M. J., \& Michels, D.
J. 1985, J. Geophys. Res., 90, 8173

\bibitem[Hu 1989]{hu89}
Hu, Y. Q. 1989, J. Comput. Phys., 84, 441

\bibitem[Hu 2005]{hu05}
Hu, Y. Q. 2005, in IAU Symp. 266, Coronal and Stellar Mass
Ejections, ed. K. Dere, J. Wang, \& Y. Yan (Cambridge: Cambridge
Univ. press), 263

\bibitem[hu 2003a]{hu03}
Hu, Y. Q., Habbal, S. R.,  Chen, Y., {\&} Li, X., 2003a, 108(A10),
1377, doi:10.1029/2002JA009776

\bibitem[Hu et al. 2003b]{hue03}
Hu, Y. Q., Li, G. Q., \& Xing X. Y. 2003b, J. Geophys. Res.,
108(A2), 1072, doi:10.1029/2002JA009419

\bibitem[Hu et al. 1997]{hux97}
Hu Y. Q., Xia L. D., Li X, Wang J. X., \& Ai G. X., 1997, Sol.
Phys., 170, 283

\bibitem[Isenberg (1993)]{isenberg93}
Isenberg, P. A., Forbes, T. G., \& Demoulin, P., 1993, ApJ, 417,
368

\bibitem[Jing (2005)]{jing05}
Jing, J., Qiu, J., Lin, J., Qu, M., Xu, Y., \& Wang, H., 2005,
ApJ, 620, 1085

\bibitem[Krall \& Chen (2000)]{krall00}
Krall, J., Chen, J., \& Santoro, R., 2000, ApJ, 539, 964

\bibitem[Li \& Hu 2003]{li03}
Li, G. Q., \& Hu, Y. Q. 2003, Chinese J. Astron. Astrophys., 3, 555

\bibitem[Lin \& Forbes 2000]{lin00}
Lin, J., \& Forbes, T. G. 2000, J. Geophys. Res., 105(A2), 2375

\bibitem[Lin (1998)]{lin98}
Lin, J. Forbes, T. G., Isenberg, P. A., \& Demoulin, P., 1998,
ApJ, 504, 1006

\bibitem[Lin et al. 2003]{lin03}
Lin, J., Soon, W., \& Baliunas, S. L. 2003, NewA Rev., 47, 53

\bibitem[Low (2001)]{loe01}
Low, B. C., 2001, J. Geophys. Res., 106, 25141

\bibitem[Peng \& Hu 2005]{pen05}
Peng, ZH., \& Hu, Y. Q. 2005, Chinese J. Space Sci., 25, 81

\bibitem[Qiu (2005)]{Qiu05}
Qiu, J., \& Yurchyshyn, V. B., 2005, ApJ, 634, L121

\bibitem[Sheeley 1999]{She99}
Sheeley, N. R., Jr., Walters, H., Wang, Y.-M., \& Howard, R. A.
1999, J. Geophys. Res., 104, 24739

\bibitem[Sun \& Hu 2000]{sun05}
Sun, S. J., \& Hu, Y. Q. 2005, J. Geophys. Res., 110, A05102,
doi:10.1029/2004JA010905

\bibitem[Sunhu 2007]{sun07}
Sun, S. J., Hu, Y. Q., \& Chen, Y., 2007, ApJ, 654, L167

\bibitem[vrs 2005]{vrs05}
Vr$\check{s}$nak, B., Sudar, D., \& Ruzdjak, D. 2005, A\&A, 435,
1149

\bibitem[yurchyshyn 2005]{yur05}
Yurchyshyn, V., Yashiro, S., Abramenko, V., Wang, H., \&
Gopalswamy, N. 2005, ApJ, 619, 599

\bibitem[Zhang \& Dere 2006]{zha04}
Zhang, J., \& Dere, K. P. 2006, ApJ, 649, 1100

\bibitem[Zhang et al. 2005]{zha05}
Zhang, Y. Z., Hu, Y. Q., \& Wang, J. X. 2005, ApJ, 626, 1096

\end{thebibliography}
\end{document}